\documentclass[twocolumn]{aastex701}

\graphicspath{{./}{figures/}}
\usepackage{graphicx}
\usepackage{gensymb}
\usepackage{footnote}

\begin{document}

\title{Discovery of giant bubbles in the hot gaseous halo of the massive disk galaxy NGC\, 6286}
\author[0000-0002-7875-9733]{Lin He}
\email[show]{helin@smail.nju.edu.cn}
\affiliation{School of Astronomy and Space Science, Nanjing University, Nanjing 210023, China}
\affiliation{Key Laboratory of Modern Astronomy and Astrophysics, Nanjing University, Nanjing 210023, China}

\author[0000-0003-0355-6437]{Zhiyuan Li}
\email[show]{lizy@nju.edu.cn}
\affiliation{School of Astronomy and Space Science, Nanjing University, Nanjing 210023, China}
\affiliation{Key Laboratory of Modern Astronomy and Astrophysics, Nanjing University, Nanjing 210023, China}
\affiliation{Institute of Science and Technology for Deep Space Exploration, Suzhou Campus, Nanjing University, Suzhou 215163, China}

\author[0000-0001-9062-8309]{Zongnan Li}
\email{zongnan.li@astro.nao.ac.jp}
\affiliation{National Astronomical Observatory of Japan, 2-21-1 Osawa, Mitaka, Tokyo, 181-8588, Japan}

\author[0000-0002-7077-308X]{Rubén García-Benito}
\email{rgb@iaa.es}
\affiliation{Instituto de Astrofśica de Andalucía (IAA/CSIC), Glorieta de la Astronomía s/n Apdo. 3004, 18080 Granada, Spain}

\author[0000-0001-9321-6000]{Yuanqi Liu}
\email{yuanqi@shao.ac.cn}
\affiliation{Shanghai Astronomical Observatory, CAS, 80 Nandan Road, Shanghai 200030, China}

\author[0000-0001-9062-8309]{Meicun Hou}
\email{houmc@nju.edu.cn}
\affiliation{Institute of Science and Technology for Deep Space Exploration, Suzhou Campus, Nanjing University, Suzhou 215163, China}

\begin{abstract}
Based on archival {\it Chandra} X-ray observation, optical integral-field spectroscopic data and radio interferometric data, we report the discovery of a pair of giant bubbles (with a projected radius $\sim 5$ kpc) of ionized gas emerging from a highly inclined starburst galaxy NGC\,6286, which is undergoing strong tidal interactions with its close neighbor NGC\,6285. The bubbles are manifested by extraplanar soft X-ray emission with an X-shaped morphology, which is tightly co-spatial with $\rm H\alpha$ line emission and partially coincident with radio continuum.
Low surface brightness diffuse X-ray emission can be traced out to $\sim$ 90 kpc from the galactic center, revealing the presence of an extended hot gaseous halo. 
X-ray spectral analysis of the bubbles yields a gas temperature of $0.70^{+0.16}_{-0.18}$ keV, a relatively high value among known galactic-scale bubbles in the local universe. 
An average energy injection rate of $\rm \sim 10^{43}~erg~s^{-1}$ is required to inflate the bubbles within an estimated dynamical age of $\sim$ 6.4 Myr.
The multi-wavelength properties of the bubbles can be understood with the conventional superwind scenario, in which disk/halo gas is swept up into an expanding, cooling shell by a hot tenuous wind. 
The current starburst in NGC\,6286 is energetically sufficient to launch the superwind, although we cannot rule out the possibility of a more violent AGN in the recent past as the driving source. Future high-resolution spectroscopic observations will help to shed light on the origin of the superwind and its role as an important galactic feedback process.

\end{abstract}

\section{Introduction} \label{sec:intro}

Galactic feedback, in the form of supernovae (SNe) and/or active galactic nuclei (AGN), is now widely believed to have paramount importance in the formation and evolution of galaxies. It regulates the global star formation and supermassive black hole (SMBH) accretion by driving high-velocity, large-scale outflows, which inject vast energy into the interstellar medium (ISM) and the circumgalactic medium (CGM), thereby heating and shaping the gaseous content of the host galaxy \citep{Veilleux_2005,Springel_2005,Dekel_2006,Hopkins_2012,Fabian_2012,Cicone_2014}. 

An intriguing and spectacular structure can be formed during this process: the galactic-scale bubble, akin to the well known Fermi bubbles \citep{Su_2010} and eROSITA bubbles \citep{Predehl_2020} observed in our own Galaxy. Recently, analogs of the Milky Way's bubbles have been identified in nearby galaxies, leveraging on multi-band observations including radio continuum, optical emission lines and X-rays \citep{Medling_2021,Zeng_2023,Li_2024,Heesen_2024}. However, the known sample of such bipolar, large-scale bubbles remains scarce in the local universe, and their physical origins are still debated, with both stellar and AGN-powered models being proposed \citep{Yang_2012,Crocker_2015,Miller_2016,Zhang_2020,Zhang_2021}. Unlike the rarity in current observations, state-of-the-art cosmological simulations predict that powerful galactic outflows can effectively produce bubble-like structures. In particular, \cite{Pillepich_2021} found in the Illustris-TNG50 simulations that 127 out of 198 Milky Way and Andromeda-like galaxies at $z=0$ exhibit large-scale bubbles, shells, or cavities in their gaseous halos, manifesting the efficient AGN feedback through episodic, kinetic, wind-like energy injections implemented in the Illustris-TNG simulations \citep{Weinberger_2018}. This raises important questions: Are the Fermi/eROSITA bubbles a common feature in other nearby disk galaxies? Is their apparent rarity due to insufficient sensitivity of current observations, or more an intrinsic phenomenon controlled by physical factors such as momentum/energy injection rates, mass loading efficiency as well as properties of the CGM? To address these questions, it is highly desired to expand the observed sample of galactic-scale bubbles. 

NGC\,6286 is a massive member (with a stellar mass $M_* \approx 1.26\times 10^{11}~M_{\odot}$, \citealp{Howell_2010}) of the early-stage merger Arp 293, located at a distance of 80.6 Mpc ($z=0.01856$), with a projected separation of $\sim$ 33 kpc from its interacting neighbor, NGC\,6285 (Figure~\ref{fig:field}a ). Due to the high inclination angle of its disk ({\it i} $\sim 76^\circ$), a conspicuous stellar feature extending southeast of the galactic plane is discernible in DESI $g$-band image (Figure~\ref{fig:field}a), reminiscent of a polar ring formed by tidal interactions (PRC C-51 in the classical polar ring catalog of \citealp{Whitmore_1990}). 
The close encounter between these two galaxies may have triggered intense starbursts, as evidenced by the high infrared luminosity of NGC\,6286 (among the so-called luminous infrared galaxies with log $L_{\rm IR}$/$L_{\odot}$ = 11.36), which suggests a star-formation rate of 41.3 $\rm M_{\odot}~yr^{-1}$ \citep{Howell_2010}. Thanks to the unprecedented sensitivity of {\it NuSTAR} at hard X-rays, a low-luminosity AGN with $L_{2-10~\rm keV} \sim (3-10)\times 10^{41}~\rm erg~s^{-1}$, obscured by a Compton-thick column density ($N_{\rm H}\gtrsim \rm 10^{24}~cm^{-2}$), is unveiled in the nucleus of NGC\,6286 \citep{Ricci_2016}. Additionally, this galaxy is also a target of the Calar Alto Legacy Integral Field Area Survey (CALIFA, \citealp{Sanchez_2012,Sanchez_2023}), and has been identified as a candidate outflow/wind galaxy due to its 
extraplanar warm ($\sim10^4$ K) ionized gas with a butterfly-like morphology (see Fig. 1 in \citealp{LC_2019}; also \citealp{Shalyapina_2004}). This extraplanar gas has been classified as shock-ionized based on various optical diagnostic diagrams \citep{Sanchez_2024}, while its exact physical origin remains unclear. 

On the other hand, the diffuse hot ($\gtrsim10^6$ K) gas of NGC\,6286 remains largely unexplored. For this reason, we have examined archival {\it Chandra} X-ray observations of NGC\,6286 (Section~\ref{sec:prep}), revealing previously unidentified extraplanar soft X-ray emission, which has an X-shape morphology and co-spatial with H$\alpha$ emission as well as radio continuum  (Section~\ref{sec:results}). 
This multi-wavelength extraplanar structure is identified as a new sample of galactic-scale bubbles and  best understood as a multi-phase outflow driven by recent nuclear activities (starburst and/or AGN) in NGC\,6286 (Section~\ref{sec:discussion}).

\section{Observations and Data preparation} \label{sec:prep}
\subsection{The {\it Chandra} data} \label{sec:Chandra}


NGC\,6286 was observed by the {\it Chandra} X-ray Observatory in 2009 (ObsID: 10566, PI: Swartz) with its Advanced CCD Imaging Spectrometer (ACIS).
The data were downloaded from the public archive and reprocessed following standard procedures, using CIAO v4.13 and the calibration files CALDB v4.9.5. 
The aim-point of the observation was placed on the S3 CCD and between NGC\,6286 and NGC\,6285. 
Only data collected by the S2 and S3 CCDs were included in the following analysis. 

We produced counts and exposure maps at a natal pixel scale of $0.492\arcsec$ in the 0.5--2 keV band. Time intervals affected by strong background flares were excluded, resulting in an effective exposure of $\sim$ 13.8 ks. The corresponding instrumental background map was generated from the stowed background files, after normalizing with the 10--12 keV count rate. Using the CIAO tool {\it wavdetect}, we detected 41 point sources across the entire field in the 0.5--2 keV band, adopting a false detection probability threshold of $10^{-6}$. For the analysis of diffuse emission, pixels within twice the 90\% encircled energy radius of each detected point source were removed uniformly from the counts, background and exposure maps. The final flux map was then created by subtracting the particle background from the total counts and dividing by the vignetted exposure.

\subsection{The CAHA PPAK data} \label{sec:CALIFA}
\begin{figure*}[htb]
\centering
\hspace{-1.cm}\includegraphics[height=8cm]{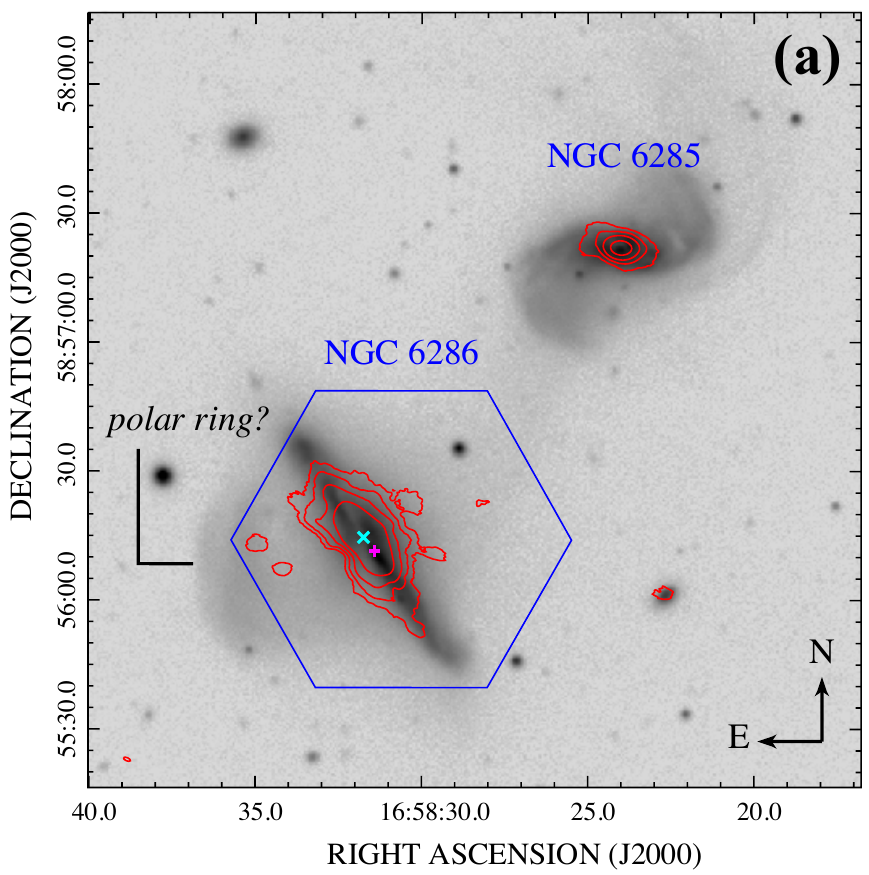}
\hspace{0.1cm}\includegraphics[height=8cm]{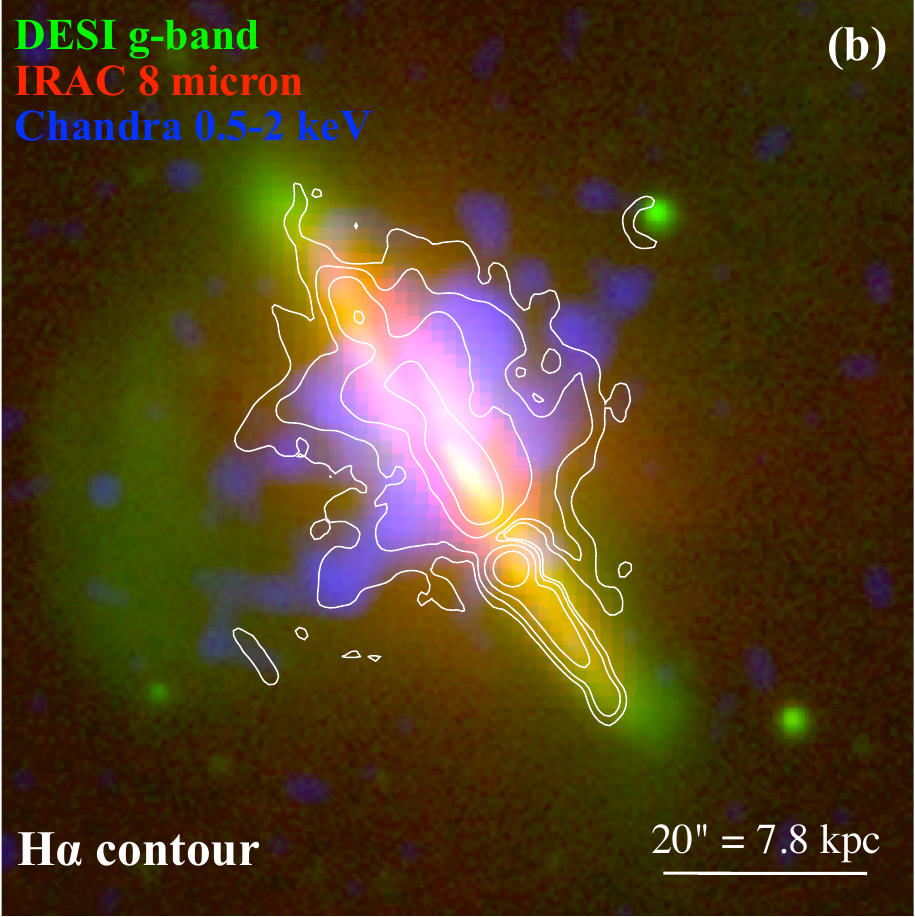}
\includegraphics[height=8cm]{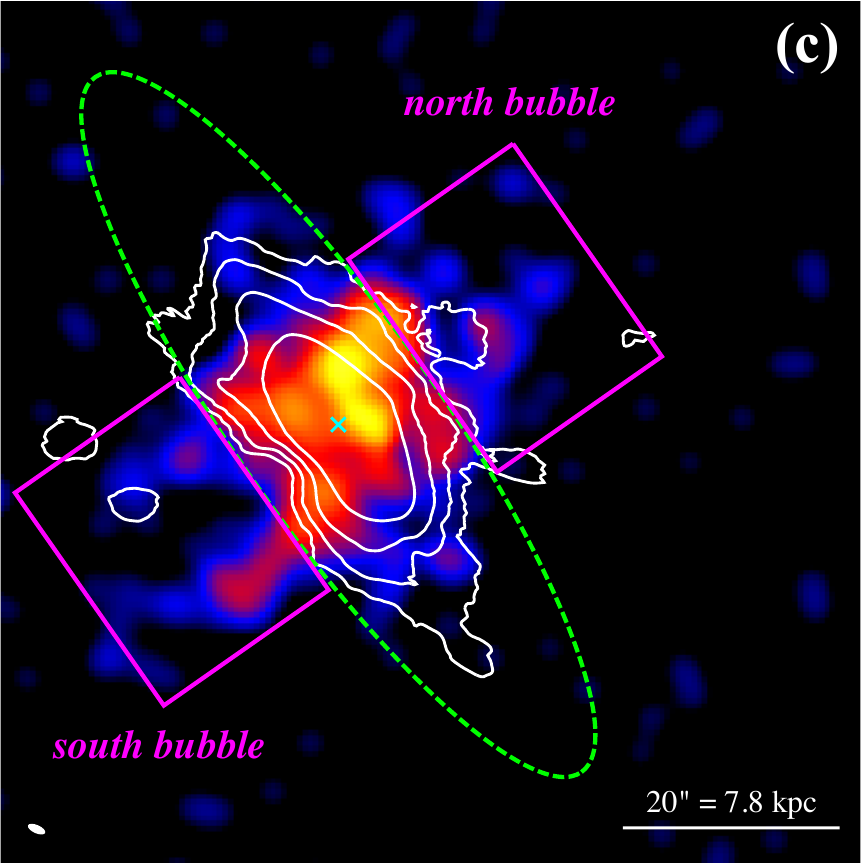}
\includegraphics[height=8cm]{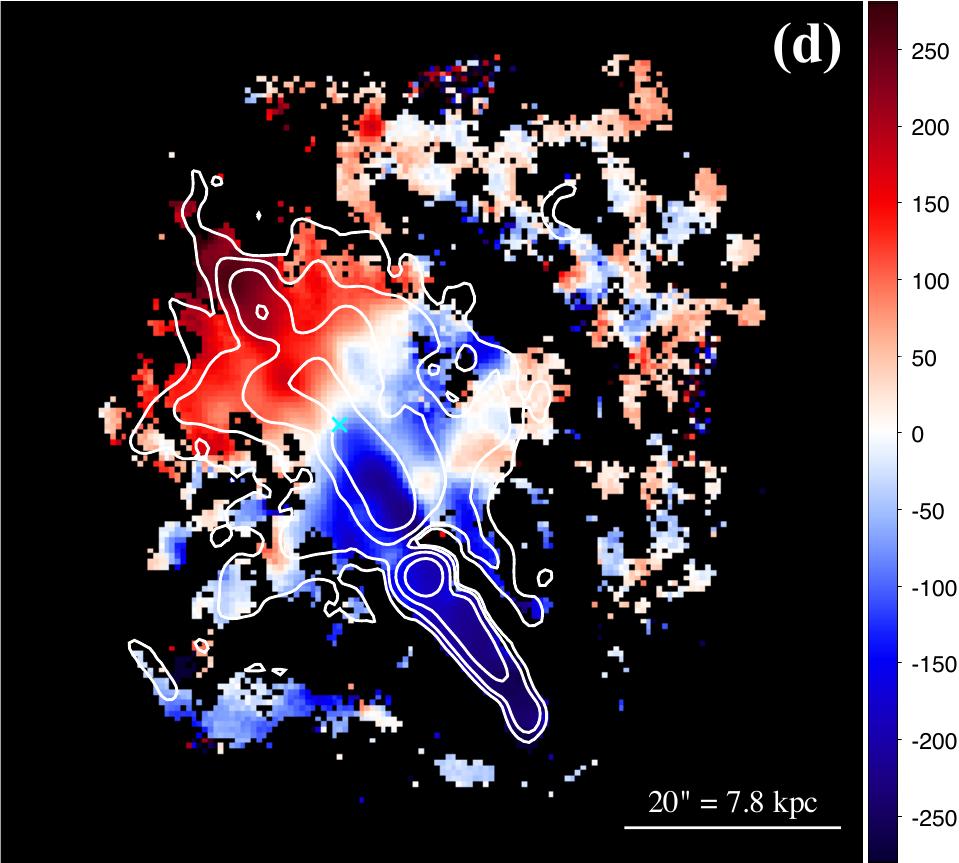}
\caption{(a): DESI $g$-band image of the interacting system Arp 293 (NGC\,6286/6285), overlaid with the VLA 3 GHz intensity contours at 5, 10, 20, 50 times the rms level (0.04 mJy/beam). The blue polygon outlines the CALIFA footprint. A prominent stellar feature located at in the southeast halo of NGC\,6286 is reminiscent of a polar ring created by tidal interaction. 
The cyan cross marks the position of the radio core, which is significantly offset from the optical center (the magenta plus). 
(b): A tri-color composite image of Spitzer/IRAC 8 $\mu$m (red), DESI $g$-band (green) and {\it Chandra} 0.5--2 keV (blue), with the $\rm H\alpha$ line emission in white contours, at intensity levels of $(1, 2.25, 6.75, 20.25) \times \rm 10^{-18}~erg~s^{-1}~cm^{-2}~arcsec^{-2}$. The optical and infrared images suggest a strongly perturbed disk, likely due to tidal interaction.
(c): {\it Chandra} 0.5--2 keV flux map zoomed-in to NGC\,6286, with point sources subtracted. A Gaussian smoothing with kernel radius of 5 pixels is applied. The green ellipse has PA$=35\degree$, $a=0.65\arcmin$ and $b=0.16\arcmin$, where PA, $a$ and $b$ are the position angle, semi-major axis and semi-minor axis, respectively, adopted from the NASA/IPAC Extragalactic Database (NED) based on the 2MASS Ks band isophote of 20.0 mag $\rm arcsec^{-2}$. The two magenta boxes outline the spectral extraction region for the X-ray bubbles. The cross and contours have the same meaning as panel (a).
(d): $\rm H\alpha$ velocity map (relative to the systemic velocity) derived from the CALIFA datacube, with the same field-of-view as panel (c). Only pixels with signal-to-noise ratio (S/N) larger than 5 are displayed. The color bar has units of $\rm km~s^{-1}$. The cross marks the position of the radio core and the contours have the same meaning as panel (b).}

\label{fig:field}
\end{figure*}

The optical spectroscopic observation of NGC\,6286 was part of the CALIFA survey, 
carried out in 2015 with the PPAK integral field unit (IFU) of the Potsdam Multi-Aperture Spectrograph instrument at the 3.5 m telescope of the Centro Astronomico Hispano Aleman (CAHA) at Calar Alto. The field-of-view is approximately a hexagon of $74'' \times 64''$ (Figure~\ref{fig:field}a), sampled by 2.7$''$ fibers. 
This field was observed with the low-resolution V500 grating, employing a three-point dithering pattern with each exposure lasting 900s.
The final wavelength coverage of the data ranges from 3745 to 7500 $\rm\AA$, with a spectral resolution of $\sim 6~\rm\AA$. The data reduction procedure follows the standard automated pipeline of CALIFA \citep{2012A&A...538A...8S, 2016A&A...594A..36S}. 
The stellar continuum in each bin was fitted using the penalized pixel-fitting method (pPXF; \citealp{2004PASP..116..138C}), with templates from the Flexible Stellar Population Synthesis (FSPS) library \citep{2009ApJ...699..486C, 2010ApJ...712..833C}.
The residual spectra, obtained after subtracting the stellar model, reveal prominent emission lines such as H$\alpha$ ($\lambda$6563), H$\beta$ ($\lambda$4861), [N\,II] $\lambda\lambda$6548, 6584, [O\,III] $\lambda\lambda$4959, 5007, and [S\,II] $\lambda\lambda$6716, 6731. These lines serve as standard tracers of warm ($\sim 10^4$ K) ionized gas. On a pixel-by-pixel basis, we derived the emission line parameters, including the line centroid (line-of-sight velocity), width (velocity dispersion) and line flux by fitting each line with a single Gaussian profile. 
In the following, we focus on the H$\alpha$ line that best traces the extraplanar ionized gas structure, but defer a detailed study of the other emission lines and the stellar component in a future work.

\subsection{The VLA data} \label{sec:Chandra}
We also utilized archival VLA observations of NGC\,6286, taken at S-band (central frequency 3 GHz) under both A- and C-configuration. The data were reduced with the Common Astronomy Software Applications (CASA; Team et al. 2022) version pipeline 6.5.4. and then applied to the VLA calibration pipeline. 
To better probe the potential extended continuum emission, we combined the calibrated A- and C-array data and adopted natural weighting for the imaging analysis. The resultant imaging has a beam size of $1.7\arcsec \times 0.8\arcsec$ and an rms level of 0.04 mJy/beam.

\section{Revealing a pair of giant X-ray bubbles} \label{sec:results}

\subsection{Diffuse X-ray morphology: an X-shaped structure} \label{sec:map}

The 0.5--2 keV flux map, shown in Figure \ref{fig:field}, reveals a previously unreported X-shaped morphology of the extraplanar diffuse X-ray emission, which is oriented perpendicular to the galactic disk (represented by the green ellipse) and reaches out to a height of $\sim 0.3\arcmin$ ($\sim$7 kpc) from the mid-plane on the southern (northern) side. The apparent asymmetry between the southern and northern sides could be partially intrinsic and partially due to extinction cast by the highly inclined disk.
Remarkably, this X-ray feature is spatially coincident with extraplanar $\rm H\alpha$ emission on both sides of the disk (demonstrated by the tri-color image in Figure \ref{fig:field}b), 
previously reported to have a butterfly-like morphology \citep{LC_2019}.
It is readily conceivable that the co-spatial X-ray and $\rm H\alpha$ emission trace the expanding shells (with a projected thickness of 1.8 kpc) of a pair of giant bubbles inflated by an energetic galactic outflow (an alternative interpretation is addressed in Section \ref{subsec:tidal}).  

Additionally, significant ($ > 5\sigma$) 3 GHz radio continuum emission, outlined by the contours in Figure \ref{fig:field}c, is also detected both in and outside the disk region. 
In particular, enhanced radio emission is present at the roots of the southern bubble, but the
moderate angular resolution of the VLA image prevents an unambiguous association with the X-ray and H$\alpha$ emission there.  
Moreover, a clump (at $5\sigma$ significance) of radio emission is seen coincident with the tip of the upper X-ray/$\rm H\alpha$ shell of the southern bubble. 
On the other hand, no radio counterpart can be clearly identified with the northern bubble. 

The line-of-sight velocity field of the $\rm H\alpha$-emitting gas is shown in Figure~\ref{fig:field}d, with the $\rm H\alpha$ intensity contours superposed.
The gas in the disk shows a clear rotation pattern, such that the receding (approaching) side is on the northeast (southwest). Notably, gas in the shells follow a similar rotation pattern, with the absolute line-of-sight velocity decreasing with increasing latitude (a more detailed analysis of the $\rm H\alpha$ kinematics is provided in Appendix \ref{sec:appendixB}, where we subtract a pure rotational disk model from it and evaluate the residuals). 
This strongly suggests that the shells consist of gas pulled out from the rotating disk, consistent with a picture in which an expanding hot wind-blown bubble sweeps up/entrain ambient cold gas (Section~\ref{subsec:superwind}).

\subsection{X-ray intensity profiles} \label{sec:profile}

\begin{figure*}[htb]
\centering
\includegraphics[width=0.49\textwidth]{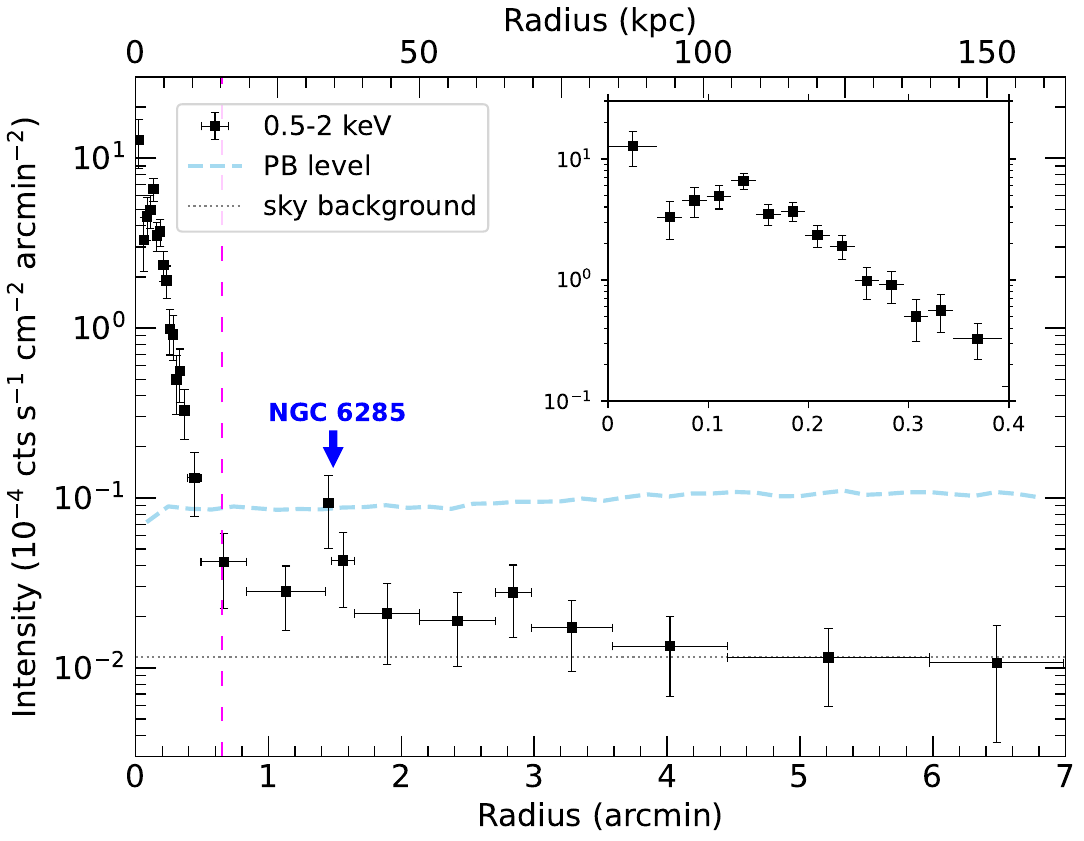}
\includegraphics[width=0.49\textwidth]{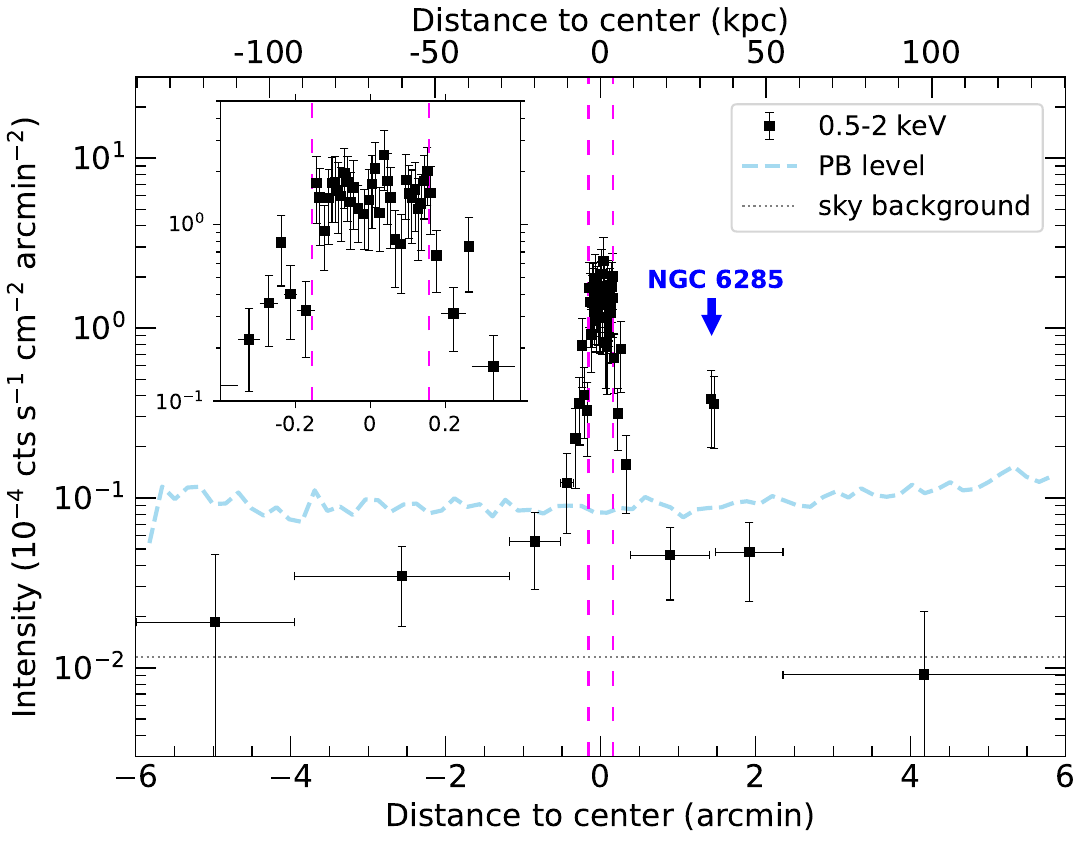}
\caption{Instrumental background-subtracted, exposure-corrected radial ({{\it left}) and vertical ({\it right}) X-ray intensity profiles of NGC\,6286 in the 0.5--2 keV band, adaptively binned to achieve a S/N greater than 2. The skyblue and grey lines indicate the level of instrumental background and sky background, respectively. The arrow indicates the position of the companion galaxy NGC\,6285. 
The vertical dashed lines indicate the size of the semi-major axis in the left panel and the semi-minor axis in the right panel, respectively.
A zoom-in view of the inner 0.4$\arcmin$ (1.6 kpc) is shown by the inserts. }
}
\label{fig:intensity}
\end{figure*}

To quantify the extent of the diffuse X-ray emission, we extracted both radial and vertical 0.5--2 keV intensity profiles for NGC\,6286, as shown in Figure \ref{fig:intensity}. By definition, the radial intensities are calculated within sequential concentric annuli, and the vertical intensities are derived from parallel slides (vertical to the minor-axis) with a length of $2a=1.3\arcmin$, 
where $a$ is the semi-major axis determined from 2MASS Ks-band image.
Both profiles have been corrected for instrumental background and vignetting, and are adaptively binned to achieve a S/N greater than 2. 
Evident in both profiles, the diffuse X-ray emission peaks toward the galactic center/mid-plane, consistent with the actively star-forming disk.
Outside the disk region,
apart from an abrupt enhancement at the radius of $\sim 1.5\arcmin$, which corresponds to the companion galaxy NGC\,6285, the diffuse X-ray emission exhibits a clear excess relative to the local background, out to a projected distance at least $\sim 4\arcmin$ ($\sim 90$ kpc) on the southern (negative) side and $\sim 2\arcmin$ ($\sim 45$ kpc) on the northern side.
Notably, this goes well beyond the height of the bubble shells (represented by the two mirrored tips located a vertical distance of $\pm 0.25\arcmin$ in the vertical profile, see the insert panel), revealing the existence of an extended hot gaseous halo in NGC\,6286. 

\subsection{X-ray spectral analysis} \label{sec:spectra}

To characterize the hot plasma residing in the bubble shells, we extracted their X-ray spectrum from the two rectangular regions marked in Figure \ref{fig:field}c. Due to the limited statistics, we consider only the summed spectrum of the southern and northern bubbles. An annular sector region on the S3 CCD, with inner and outer radii of $4\arcmin$ and $7\arcmin$, is adopted as the local background. The spectrum was adaptively binnned to achieve a S/N greater than 2 
and fitted with Xspec v12.14.0. 
A single thermal plasma model (APEC in Xspec) is found to fit the spectrum reasonably well ($\chi^2\rm / dof=16.39/13$), which gives a best-fit gas temperature of $0.70^{+0.16}_{-0.18}$ keV. Metallicity was poorly constrained and thus fixed at the solar value. The APEC normalization is $\rm 1.55\times 10^{-5}~cm^{-5}$, corresponding to a gas density of $\rm \sim 0.02~cm^{-3}$, assuming a spherical shell volume with inner and outer radius of 2.9 and 4.7 kpc. The estimated enclosed hot gas mass is $1.2f_h^{0.5}\times 10^8~M_{\odot}$ (where $f_h$ is the volume filling factor), with a 0.5--2 keV luminosity of $\rm (3.5\pm 0.4)\times 10^{40}~erg~s^{-1}$. 
For comparison, the mass of the warm ionized gas enclosed in the shells is estimated to be $3f_w^{0.5}\times 10^8~M_{\odot}$ (where $f_w$ is the volume filling factor), based on the observed H$\alpha$ luminosity ($\rm 1.6\times 10^{40}~erg~s^{-1}$) and assuming the standard case B emissivity.

\begin{figure*}[htbp]
\centering
\includegraphics[width=0.4\textwidth]{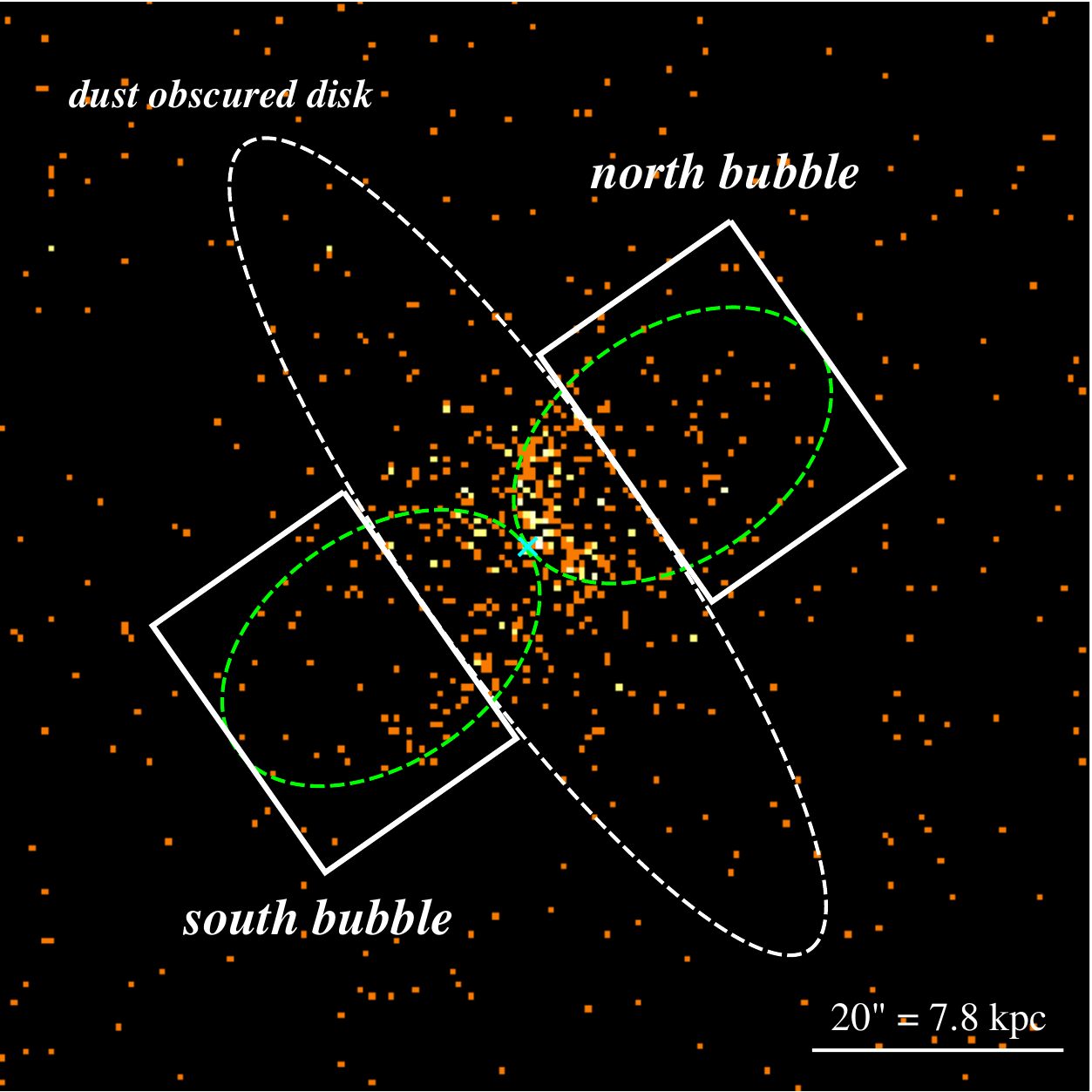}
\includegraphics[width=0.55\textwidth]{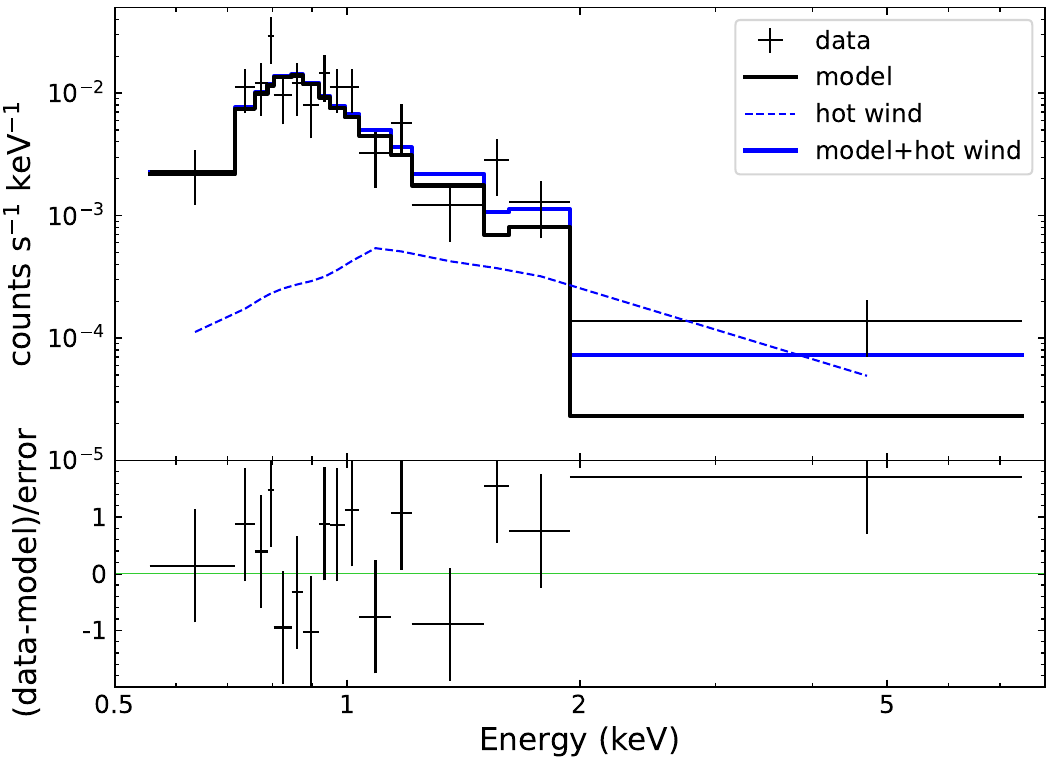}
\caption{
{\it Left}: 0.5--8 keV counts image centered on NGC\,6286. The white ellipse and boxes demarcate the disk and bubble emission, respectively, while the two green ellipses are used to outline the shells of the putative bipolar bubbles.
{\it Right}: Observed X-ray spectrum of the bubbles, adaptively binned to achieve a S/N ratio of 2 per bin. The best-fit model (black solid line), $tbabs\times apec$, represents the shells dominating the observed soft X-ray emission.  The lower panel plots the relative residual. A putative hot wind component, i.e., a second apec with a fixed temperature of $kT=$ 5 keV, is also plotted (blue dash line) for illustration.  
}
\label{fig:spectrum}
\end{figure*}

Considering the superwind scenario (Section \ref{subsec:superwind}), it is tempting to place constraints on the putative hard X-ray emission from the very hot superwind. To do so, we introduce an additional {\it apec} to the spectral model, fixing its temperature at 5 keV and requiring its normalization to roughly match the observed spectral bin above 2 keV (Figure~\ref{fig:spectrum}). In this case, the superwind component has a 0.5--8 keV luminosity of $\rm 6.5\times 10^{39}~erg~s^{-1}$, an estimated gas density of 0.017 cm$^{-3}$ and thermal energy of $\rm 8\times 10^{56}$ erg.

Albeit strongly desired, a meaningful spectral analysis of the extended halo, which is farther from the disk, is not feasible due to the limited statistics. Future deep observations focusing on the soft X-rays are required to better quantify the hot gas halo of this galaxy.

\section{Summary and discussion} \label{sec:discussion}
Based on multi-wavelength data from {\it Chandra}/ACIS, CAHA/PPAK and VLA, we have revealed a pair of giant bubbles emerging from the disk of NGC\,6286, a highly inclined starburst galaxy in an on-going merger. 
The bubbles are manifested by extraplanar X-ray emission with a prominent X-shaped morphology, consisting of mirrored shells reaching out to $\sim$7 kpc on both sides of the disk, which are tightly co-spatial with warm ionized gas traced by H$\alpha$ emission and are partially coincident with radio continuum. The bubbles are themselves embedded in an extended hot gas halo, which is traced by a low surface brightness diffuse X-ray emission detectable out to a projected distance of $\sim$90 kpc from the galactic center.  
Here we discuss the possible origin of the giant bubbles in NGC\,6286 and compare them with other known galactic-scale bubbles in the local universe.

\subsection{A tidal feature?}
\label{subsec:tidal}
While it is tempting to interpret the observed bubbles as the result of an energetic outflow driven by a starburst and/or an AGN, 
the fact that NGC\,6286 is under strong interaction with its close neighbor, NGC\,6285, suggests the possibility of tidal force to pull out stars and gas from the disk. 
Indeed, the ``polar ring''-like stellar feature (Figure \ref{fig:field}a; \citealp{Shalyapina_2004}) can be understood as a tidal relic, and the morphologically perturbed stellar disk of NGC\,6286 (Figure \ref{fig:field}b) is also evidence of tidal perturbation. 
However, 
we argue that the observed shells consisting of hot and warm gas is unlikely to be a tidal feature for the following reasons: (i) It is hardly conceivable that tidally stripped gas can simultaneously result in spatial coincidence between the $\rm H\alpha$ and X-ray emission, and a natural alignment with the minor axis, as illustrated in Figure~\ref{fig:bubble}.
(ii) The mass of the companion galaxy is only $\sim 10\%$ of NGC\,6286 (estimated from their rotation curves, \citealp{Shalyapina_2004}), and the tidal force it exerts on the inner region of NGC\,6286 
is probably inadequate to counteract the gravitational pull of NGC\,6286 itself,
even if the two galaxies were once closer to each other. 
Specifically, the tidal radius, at which stars and gas within a galaxy become unbound or stripped, is given by $D[\frac{m}{M(3+e)}]^{1/3}$ \citep{Read_2006}, where {\it m} and {\it M} are the masses of the two interacting galaxies, {\it D} is their separation and {\it e} is the orbital ellipticity. Assuming a mass ratio of 10, a (projected) separation of 33 kpc, and a circular orbit, the tidal radius of NGC\,6286 is $\sim$ 45 kpc, much larger than the physical size of the X-ray structure ($\sim$7 kpc), suggesting the peculiar structure is unlikely to be tidal stripped gas. 
(iii) Tidal forces alone cannot heat the stripped gas to the X-ray-emitting temperatures. Tidal interactions typically involve shearing effects and shock-heating. While the former should not increase the gas temperature, the latter, predicted by simulations, can thermalize ambient gas to $\sim 10^6$ K \citep{Cox_2006,Sinha_2009}. The after-shock temperature is given by $T_{\rm final}=\frac{3\mu m_{\rm H}}{16k}v_{\rm initial}^2$ \citep{Sinha_2009}, where $T_{\rm final}$ is the final gas temperature, $\mu$ is the mean molecular weight, $m_{\rm H}$ is the proton mass, {\it k} is the Boltzmann constant and $v_{\rm initial}$ is approximately equal to the radial velocity prior to the collision. By assuming a typical collision velocity of 300 km~s$^{-1}$, the final temperature can reach $\sim0.1$ keV, which is still significantly lower than the measured value (0.7 keV).

On the other hand, X-ray emission consistent with $\rm H\alpha$ structures is common-place in classical starburst winds as seen in M82 and NGC\,253. It is conceivable, though, tidal interaction by NGC\,6285 somehow thickens the gaseous disk of NGC\,6286 and facilitates the breakout of the galactic wind to form the large-scale bubbles.

\subsection{The swept-up shell of disk/halo gas by a superwind} 
\label{subsec:superwind}
The X-shaped X-ray/H$\alpha$ emission is consistent with the conventional model of swept-up shells in starburst galaxies, proposed by \cite{Strickland_2002} to explain the spatially correlated X-ray and $\rm H\alpha$ emission in the halo of NGC\,253. In this scenario, the superwind launched by  energetic stellar winds and SNe is very hot ($10^{7-8}$ K), tenuous and weakly X-ray-emitting. The expanding wind eventually sweeps up the ambient gas (typically cold, from a thick disk or the halo) and becomes bounded by a reverse shock, where the swept-up gas is shock-heated and subsequently cool to $T\lesssim 10^4$ K, emitting $\rm H\alpha$. The shocked wind at the interface between the cool shell and the unshocked free wind gives rise to soft X-rays (with a temperature of $10^{6-7}$ K). In this picture, the soft X-ray emission is expected to locate somewhat interior to the H$\alpha$ emission. A close look at Figure~\ref{fig:field}b suggests that this is likely the case in the upper shell of the south bubble, although we cannot fully rule out the effect of self-absorption in the shell with current data.

This superwind-blown bubble scenario is further supported by the rotation pattern of the H$\alpha$ shell described in Section~\ref{sec:map}.
We speculate that the tidal interaction between NGC\,6286 and NGC\,6285 helps to form a puffed up disk, as hinted by the optical and infrared images (Figure \ref{fig:field}b), which in turn facilitates the bubble sweeping and the creating of the shells.

\subsection{Powering source of the superwind} 

The next question is what physical mechanism powers the putative superwind.
In view of the active star formation in the inner disk of NGC\,6286, it is appealing to consider the conventional scenario in which core-collapse supernovae (CCSNe) launch the superwind \citep[e.g.][]{Strickland_2002}.
Following \cite{Miller_2016}, which modelled the Fermi bubbles as a continuous outflow with the self-similar Sedov–Taylor solution \citep{Weaver_1977,ML_1988,Veilleux_2005}, the mechanical energy injection rate is $\xi \dot{E} \approx 10^{43}~\rm erg~s^{-1}\times (\frac{\it n_{\rm 0}}{10^{-2}~\rm cm^{-3}})(\frac{\it r}{4.7~\rm kpc})^2(\frac{{\it v}_{\rm exp}}{\rm 430~km^{-1}})^3$, where $n_0$ is the characteristic gas density of the thick disk/halo, $r$ is the current bubble radius, and $v_{\rm exp}$ is the expanding velocity (approximated by the sound speed of the hot gas in the shell with $kT=$ 0.7 keV).  
The dynamic age of the bubbles is $t_{\rm dyn} \approx 6.4~\rm Myr~(\frac{\it r}{\rm 4.7~kpc})(\frac{{\it v}_{\rm exp}}{\rm 430~km~s^{-1}})^{-1}$.
The total energy injected within the bubble dynamic age ($\xi \dot{E}t_{\rm dyn} \sim 2\times10^{57}$ erg) is comparable to the estimated thermal energy of the hot bubble (Section \ref{sec:spectra}). 

To account for the above inferred energy injection rate, 
an average birth rate of $0.3\xi^{-1}$ CCSN $\rm yr^{-1}$ during the dynamic age of the bubble is required.
This translates to a star formation rate (SFR) of 30$\rm \xi^{-1}~ M_{\odot}~yr^{-1}$ (assuming 0.01 CCSN per unit SFR, e.g., \citealp{Strolger_2015}). The total (obscured and unobscured) SFR of NGC\,6286, derived from the sum of ultraviolet and infrared emission, is 41.3 $\rm M_{\odot}~yr^{-1}$ \citep{Howell_2010}. 
Considering that much of this SFR takes place in the inner galaxy,
and that the CCSNe energy conversion efficiency is typically 0.1--1 (e.g., \citealp{Wada_2001,Melioli_2004,Zhang_2014}),
it seems plausible that current star formation in NGC\,6286 is energetically sufficient to launch the superwind if $\xi$ is close to 1. 
Otherwise, a more conservative efficiency would require more intense star formation to provide a reasonable power source. This can be mitigated by assuming a lower $n_0$ or $v_{\rm exp}$, which leads to less input energy and a longer dynamical age, or by incorporating the AGN as an additional power source.
Furthermore, we can infer a mass-loading efficiency $\eta \approx (M_{\rm sh}/t_{\rm dyn})/SFR \approx 1.6$, where $M_{\rm sh}$ is the total (hot plus warm) gas mass enclosed in the shell, assuming a unity volume filling factor (Section \ref{sec:spectra}). A smaller filling factor or an addition of (unseen) cold gas can substantially modify this estimate.

We shall also consider the possibility of an AGN-driven superwind. The current AGN bolometric luminosity, $\rm (7-40)\times 10^{42}~erg~s^{-1}$ \citep{Ricci_2016}, is rather moderate and unlikely to account for the inferred kinetic energy injection rate, as AGNs typically have $\xi \lesssim 0.01$ \citep{Santoro_2020}. 
However, we cannot rule out a much more violent AGN in the recent past. 
Future high-resolution observations, e.g. afforded by JWST, penetrating the starbursting nucleus, will facilitate a direct mapping of the nuclear outflow launching site, providing conclusive evidence of its origin.

\subsection{Comparison with other galactic-scale bubbles}
A few galactic-scale bubbles have previously been identified in nearby galaxies based on multi-wavelength observations.
We compare the physical size, X-ray-emitting plasma temperature and other properties between them and the newly identified bubbles in NGC\,6286 (Table~\ref{tab:comparison}). 

The NGC\,6286 bubbles have a relatively high temperature (0.7 keV), a moderate size (7 kpc) and a comparable estimated power ($\rm \sim 10^{43}~erg~s^{-1}$). A distinctive feature of the NGC\,6286 bubbles is their X-shaped morphology, which have an open top (i.e., without a `cap' observed in some other cases). Similar X-shaped X-ray emission was recently discovered in a supermassive S0 galaxy NGC\,5084, spanning $\sim$ 17 kpc, while the AGN- or starburst-powered origin cannot be discriminated \citep{Borlaff_2024}. Extended $\rm H\alpha$ nebulae with butterfly/hourglass/bubble-like structures and co-spatial X-ray emission have also been observed in the late-stage gas-rich merger NGC\,6240, spanning a few to tens of kpc, which are linked to multiple epochs of starburst driven winds \citep{Yoshida_2016,Medling_2021}. This system is further complicated by the presence of two X-ray-confirmed AGNs \citep{Komossa_2003,Max_2007}, suggesting a probable hybrid contribution from intense star formation and AGN activities. 
We do not include these two systems in Table\ref{tab:comparison} because the temperatures of the X-ray-emitting bubble plasma have not been specifically studied.  

In addition, diffuse X-ray emission associated with bubble-like structures emitting in optical/radio bands has been observed in NGC\,2782 \citep{Jogee_1998,BG_2017} and NGC\,6764 \citep{Croston_2008}, even though the morphologies do not resemble well-defined X-ray bubbles. Another case is NGC\,5775, where a shell-like feature in southwest and a blob in northeast have been identified in soft X-rays, likely representing a pair of outflowing wind bubbles partially confined by strong magnetic field \citep{Li_2008}. We regard these systems as putative X-ray bubbles and include them in Table\ref{tab:comparison} to provide a more complete census of galactic-scale bubbles. However, we treat the ultraluminous infrared galaxies (ULIRGs) differently due to their extreme nature, marked by exceptionally violent AGN and star formation activities. Extended X-ray and optical line emission nebulae, spanning tens of kiloparsecs, have been found in ULIRGs including Mrk 231, Mrk 273 and SDSS J1356+1026 \citep{Veilleux_2014,Liu_2019,Greene_2014}. They can be dominated either by AGN photoionization or shocks induced by powerful outflows. As such, these systems may not be directly comparable to other galactic bubbles discussed above.

\vskip0.5cm
To conclude, NGC\,6286 represents a rare host of galactic-scale bubbles evolving within its extended hot CGM. 
The origin of these bubbles, likely their cousins found in other nearby galaxies, remains uncertain, despite extensive multi-wavelength data. 
Future high-resolution spectroscopic observations and tailored numerical simulations will help to shed light on the formation and evolution of the bubbles, as well as the underlying feedback processes in this and other intriguing systems.

\startlongtable
\begin{deluxetable*}{ccccccccc}
\tabletypesize{\scriptsize}
\tablecaption{Comparison of Galactic-scale Bubbles \label{tab:comparison}}
\tablewidth{0pt}
\tablehead{
\colhead{Host} & \colhead{$kT$} & \colhead{$kT_1$} & \colhead{$kT_2$} & \colhead{Size} & \colhead{$t_{\rm dyn}$} & \colhead{$\xi \dot{E}$} & \colhead{Observed bands} & \colhead{Ref}
}
\colnumbers
\startdata
Milky Way  & $\sim 0.3$ & $0.21\pm 0.01$ & $0.74\pm 0.02$ & 10--15 & 5--50 & $5\times 10^{40}$--$7\times 10^{42}$ & $\gamma$-ray, X-ray & a--g \\
M106  & $0.53^{+0.03}_{-0.01}$ & $0.25^{+0.03}_{-0.02}$ & $0.76^{+0.04}_{-0.03}$ & $\sim 8$ & 4--8 & $\sim 4\times 10^{42}$ & X-ray, 144 MHz & h \\
NGC\,1482  & $0.61\pm 0.03$ & -- & -- & $\sim 1.5$ & $\sim 7.5$ & 0.4--4$\times 10^{42}$ & X-ray, $\rm H\alpha$ &  i-k \\
NGC\,3079  & $0.88^{+0.03}_{-0.04}$ & -- & -- & $\sim 1.5$ & 0.4--0.8 & 1.2--2.5$\times 10^{42}$ & X-ray, $\rm H\alpha$, 3 GHz &  l \\
NGC\,6286  & $0.70^{+0.16}_{-0.18}$ & -- & -- & $\sim 7$ & $\sim 6.4$ & $\sim 10^{43}$ & X-ray, $\rm H\alpha$, [O\,III] &  this work$^{\dagger}$ \\
\hline
NGC\,6764$^{\dagger\dagger}$  & $0.75\pm 0.05$ & -- & -- & $\sim 0.7$ & -- & $\sim 10^{42}$ & X-ray, 1.4 GHz &  m \\
NGC\,2782  & $0.69\pm 0.09$ & -- & -- & $\sim 1$ & $\sim 4$ & $\sim 1.6\times 10^{42}$ & X-ray, $\rm H\alpha$, [O\,III], 5 GHz &  n, o \\
NGC\,5775  & -- & 0.17 & $0.57\pm 0.15$ & $\sim 10$ & $\gtrsim 25$ & $\lesssim 3.8\times 10^{42}$ & X-ray &  p \\
\enddata
\tablecomments{(1) Host galaxy of the bubble; (2) Temperature of the hot plasma in units of keV, derived from X-ray spectra fitted with single-temperature model; (3)$\&$(4)  Temperature of the hot plasma in units of keV, derived from X-ray spectra fitted with two-temperature model; (5) Bubble size (diameter or height from the disk plane), in units of kpc; (6) Estimated dynamical age in units of Myr; (7) Estimated energy injection rate in units of $\rm erg~s^{-1}$; (8) Observed wavebands of the bubble; (9) References: a: \cite{Su_2010}; b: \cite{Sarkar_2015}; c: \cite{Miller_2016}; d: \cite{Predehl_2020}; e: \cite{Zhang_2020}; f: \cite{Yang_2022}; g: \cite{Gupta_2023}; h: \cite{Zeng_2023}; i: \citep{Veilleux_2002}; j: \cite{Hota_2005}; k: \cite{Vagshette_2012} l: \cite{Li_2024}; m: \cite{Croston_2008}; n: \cite{Jogee_1998}; o: \cite{BG_2017}; p: \cite{Li_2008}. \\
$\dagger$: We focus on the X-ray/$\rm H\alpha$ bubbles within the scope of this work, although the X-shaped morphology is also significant in [O\,III]. \\
$\dagger\dagger$: We regard the bubbles hosted by NGC\,6764, NGC\,2782 and NGC\,5775 as putative in virtue of their less identifiable bubble-like morphologies on the visual inspection.
}
\end{deluxetable*}

\section*{Acknowledgements}
This work is supported by the National Key Research and Development Program of China (No. 2022YFF0503402), the National Natural Science Foundation of China (grant 12225302, grant 12203001), the CNSA program D050102, and the China Manned Space Program through its Space Application System. Z.N.L. acknowledges support from the China National Postdoctoral Program for Innovation Talents (grant BX20220301) and the East Asian Core Observatories Association Fellowship. RGB acknowledges financial support from the Severo
Ochoa grant CEX2021-001131-S funded by MCIN/AEI/10.13039/501100011033 and PID2022-141755NB-I00.

\appendix

\section{Symmetry of the X-ray/$\rm H\alpha$ bubble}\label{sec:appendixA}

The bubbles we discovered in NGC\,6286 exhibit clear asymmetry between the northwest and southeast lobes, however, their X-shape morphology is noteworthy. To better illustrate this, we extract surface brightness profiles from two rectangular regions enclosing most of the bubble emission, for both the $\rm H\alpha$ and X-ray flux maps.

\begin{figure*}[htb]
\centering
\includegraphics[width=1.\textwidth]{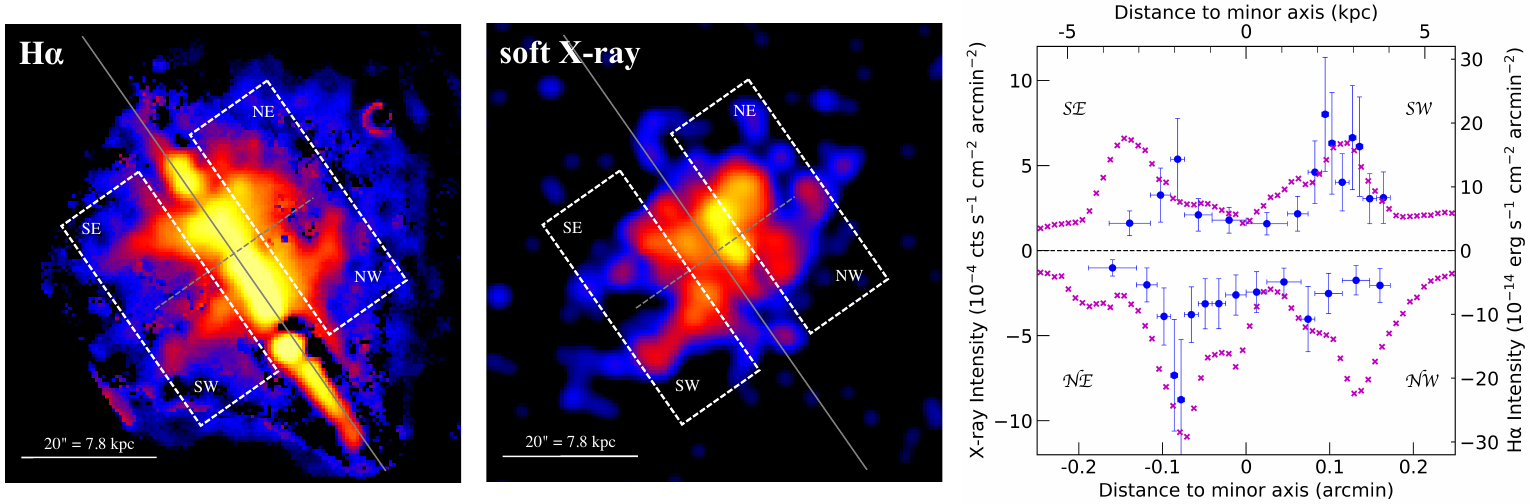}
\caption{$\rm H\alpha$ ({\it left}) and X-ray ({\it middle}) intensity maps zoomed to the bubble-like structures. The white rectangles parallel to the major axis has size of $36\arcsec\times 14\arcsec$, outlining the regions we extract the profiles. The grey solid (dashed) line indicates the major (minor) axis adopted from NED. {\it Right}: $\rm H\alpha$ (magenta) and X-ray (blue) surface brightness profiles of the bubble emission. The northern profile is defined to have negative value to show the mirror symmetry. Four peaks corresponding to the bubble boundaries are unambiguous for the $\rm H\alpha$ emission, while their X-ray counterparts are less clear especially in the north bubble.}
\label{fig:bubble}
\end{figure*}

As shown in Figure~\ref{fig:bubble}, the $\rm H\alpha$ profile clearly reveals a mirror-symmetry pattern, with four peaks unambiguously corresponding to the bubble shells in the four quadrants. A similar pattern is also observed in the X-ray profile, although the bubble shells are less distinct, particularly in the northern region. More interestingly, there is a hint that the X-ray shell emission is slightly interior to its $\rm H\alpha$ counterpart in some directions, which is consistent with the conventional model of swept-up shells proposed by \citet{Strickland_2002}. We further note that perfect symmetry of the wind/bubble structures is not expected in this highly disturbed system, as the rapid evolution and violence of the environment may distort it very quickly, similar to the case of NGC\,4676A \citep{Read_2003}.

\section{Subtraction of the standard rotating-disk model}\label{sec:appendixB}

\begin{figure*}[htb]
\centering
\includegraphics[width=1.\textwidth]{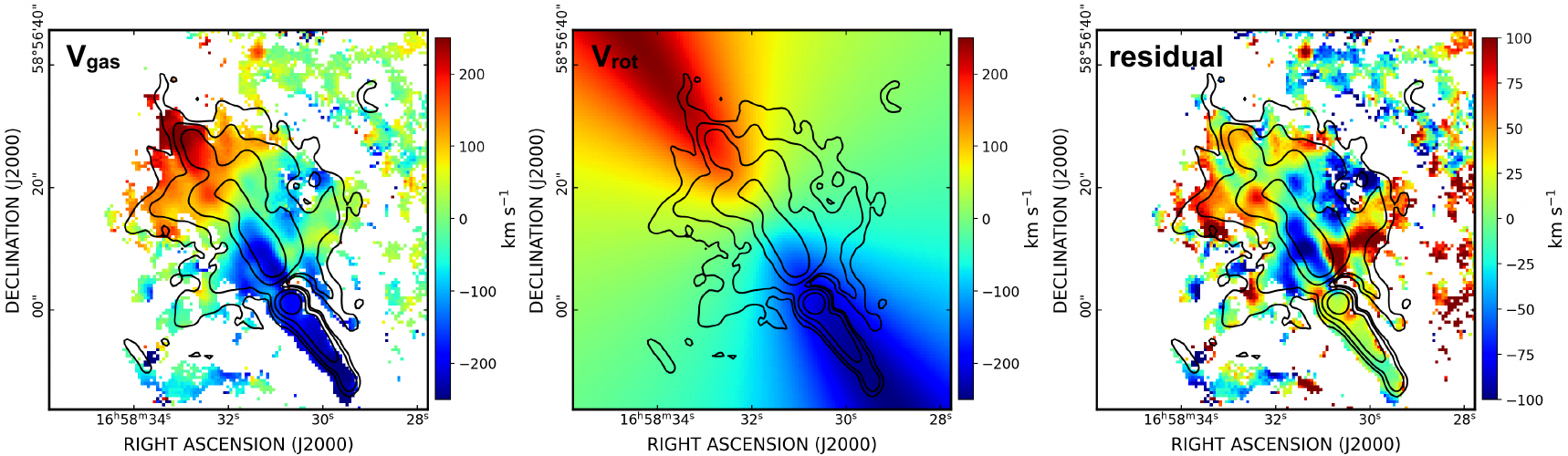}
\caption{Observed line-of-sight velocity field of the $\rm H\alpha$ gas (left), a fitted pure rotating-disk model (middle), and the residuals (right). The black contours are the same as those in Figure \ref{fig:field}.}
\label{fig:residual}
\end{figure*}

A standardized way to highlight the outflow components is to subtract the galactic disk component from the observed gas velocity field. Here we subtract a pure rotating-disk model from the $\rm H\alpha$ velocity map in Figure \ref{fig:field}, and present the residual map in Figure \ref{fig:residual}. The disk model is expressed as $V_{\rm rot}(R)=V_{\rm c}(R)~\rm sin\,{\it i}~cos\,{\psi}$, where $V_{\rm c}(R)$ is the $\rm H\alpha$ rotation velocity at radius R along the major axis, {\it i} is the inclination angle (fixed at $76^{\circ}$), {\it $\psi$} is the azimuthal angle measured from the major-axis in the galactic plane. It is shown that no clear blue-/red-shifted trend can be associated with the bubble shells after subtracting the rotational disk model. However, this is not surprising, as the bubble shells are viewed nearly edge-on, minimizing any bulk expanding motion projected into the line-of-sight. On the other hand, it is also not unreasonable to observe the original redshift/blueshift pattern in the bubble shells, because in the canonical picture, the ionized gas is entrained by the hot wind and should preserve some angular momentum of the disk where the gas was originated. Since the residual map does not provide more conclusive evidence for the putative wind-blown bubble, we leave it in the Appendix to give additional information about the $\rm H\alpha$ kinematics.

\bibliography{ref}{}
\bibliographystyle{aasjournal}

\end{document}